\documentclass[aip,apl]{revtex4-1}

\usepackage{amsmath}
\usepackage{graphicx}
\usepackage{siunitx}

\begin{document}

\title{Calibration-based overlay sensing with minimal-footprint targets}

\author{Tom A. W. Wolterink}
\author{Robin D. Buijs}
\affiliation{Center for Nanophotonics, AMOLF, Science Park 104, 1098XG Amsterdam, The Netherlands}
\author{Giampiero Gerini}
\affiliation{Optics Department, Netherlands Organization for Applied Scientific Research (TNO), Stieltjesweg 1, 2628CK Delft, The Netherlands}
\affiliation{Department of Electrical Engineering, Technische Universiteit Eindhoven (TU/e), 5600 MB Eindhoven, The Netherlands}
\author{Ewold Verhagen}
\author{A. Femius Koenderink}
\email{f.koenderink@amolf.nl}
\affiliation{Center for Nanophotonics, AMOLF, Science Park 104, 1098XG Amsterdam, The Netherlands}

\date{\today}

\begin{abstract}
Overlay measurements are a critical part of modern semiconductor fabrication, but overlay targets have not scaled down in the way devices have. In this work, we produce overlay targets with very small footprint, consisting of just a few scattering nanoparticles in two separate device layers. Using moir\'e patterns to deterministically generate many overlay errors on a single chip, we demonstrate successful readout of the relative displacement between the two layers and show that calibration on one realization of the targets can be used for overlay measurements on subsequent instances.
Our results suggest using greater quantities of smaller overlay targets may benefit performance both directly and through finer sampling of deformation.
\end{abstract}


\maketitle 

\section{Introduction}
Overlay error --- the misalignment between different patterned layers on a chip --- is a key quantity in wafer metrology \cite{Silver1995, Bunday2016}. With feature sizes shrinking and the semiconductor industry increasingly focused on few-nanometer nodes, minimizing overlay error is crucial to manufacturing yield and device performance \cite{Felix2011, Bunday2018, Orji2018}.
Typical overlay targets consist of a set of one-dimensional gratings \cite{denBoef2016}. Scatterometry on a full set of these gratings allows reconstruction of two-dimensional overlay error, based on understanding of the scattering properties versus position error \cite{Bishop1991, Bouwhuis1979}. These overlay targets are sized based on their scattering physics and do not scale down with device feature size, leaving overlay targets $20-\SI{100}{\micro\metre}$ squared in the midst of devices fabricated with nanometer-scale precision.
Moreover, the few-nanometer overlay error tolerance in advanced nodes means that non-uniform overlay across a complete wafer becomes highly relevant. Sampling overlay error at just a few positions leaves appreciable uncertainty in the wafer deformation model that predicts overlay at intermediate positions \cite{denBoef2016, Bunday2018}. It could thus be advantageous to implement smaller overlay sensors, which might be placed in more positions to better sample overlay error across the wafer. 

In this work, we perform overlay error retrieval with targets of minimal footprint: just four nanoparticles, for a target around $400 \times 400$ \si{\nano\metre\squared} in size. Using principal component analysis on angle-resolved intensity patterns, we measure overlay retrieval performance. We investigate the effect of fabrication errors on the reliability of this method by reusing calibration data for overlay measurement on a different device and comparing overlay retrieval error.

\section{Results}
\begin{figure}[htb!]\centering
\includegraphics[width=\textwidth]{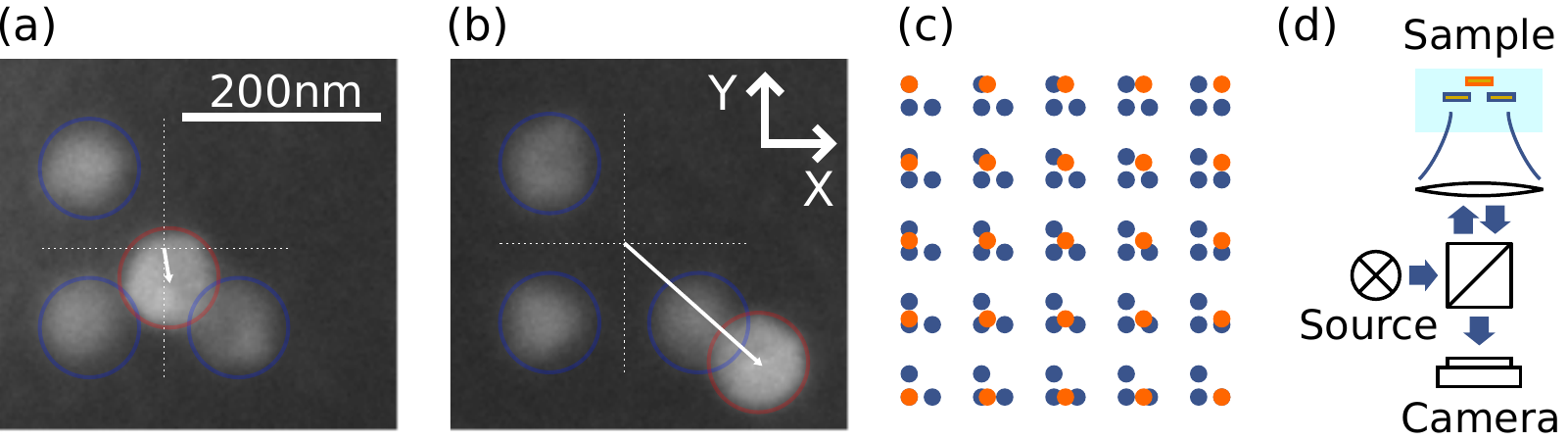}

\caption{(a) Scanning electron image of a minimal-footprint overlay target with small overlay error. Particles circled in blue sit in the bottom (far) layer, the particle circled in red sits in the upper (near) layer. (b) Like (a), but with much larger overlay error. (c) Deterministically producing many different overlay values on a single chip using the moir\'e effect. Distances in this cartoon are not shown to scale: the distance between individual targets is \SI{5}{\micro\metre} compared to the \SI{400}{\nano\metre} width of the individual targets. (d) Basic set-up for measurement of overlay error using angle-resolved intensity data. } \label{fig:method}
\end{figure}

The overlay targets we study consist of plasmonic nanoparticles. Such particles are known to couple strongly to light despite their small size \cite{vanDijk2006, Weber2004}, with scattering cross-sections larger than their geometric cross-section. This helps to maximize scattered signal despite the small area. Such particles may be created during patterning steps using metal, but they may otherwise be replaced by resonant particles made out of dielectrics \cite{Kivshar2018}. We use disc-shaped gold nanoparticles of diameter \SI{110}{\nano\metre} and height \SI{40}{\nano\metre}, distributed over two layers: three on corners of a \SI{150}{\nano\metre} square in the bottom layer and another centered on the same square, but in the top layer. The layers are fabricated on a boro-silicate glass substrate. We use electron beam physical vapor deposition to deposit gold over a shadow mask patterned by electron beam lithography. The hard mask is lifted off as a sacrificial layer, leaving the bottom layer of nanoparticles. We apply a spacer layer of \SI{89}{nm} of commercial spin-on glass (Microresist OrmoComp) and repeat the process to fabricate the top nanoparticle layer. The top layer is coated with an identical layer of spin-on glass to avoid reflections near the focal plane.
This four-nanoparticle design can show significant chiral optical response and has been proposed as a plasmonic `ruler' as the three-dimensional arrangement strongly modifies optical spectra \cite{Hentschel2012}. Such spectral tuning through spatial configuration has been demonstrated in arrays of a larger-scale variant of these structures in the infrared \cite{Hentschel2012}. The four-fold asymmetry in the design also makes the design polarization-sensitive. Two realizations of these overlay targets are shown in figure \ref{fig:method}a and \ref{fig:method}b, as imaged by scanning electron microscopy. As overlay error in this target is the distance between the center of the circumscribed square of the three bottom particles and the center position of the top particle, we see that the target in figure \ref{fig:method}a has moderate overlay error (\SI{30}{\nano\metre}) and the target in figure \ref{fig:method} has large overlay error (\SI{180}{\nano\metre}). We recognize the individual nanoparticles, imaged through the spin-on glass layer. Our goal is then to reliably and efficiently measure the overlay error in these structures.

In order to test these overlay targets over a range of overlay errors, we exploit moir\'e patterns. We use electron beam lithography to fabricate arrays of both bottom and top components of the target, but with slightly different pitch: \SI{5.0000}{\micro\metre} for the bottom layer and \SI{5.0375}{\micro\metre} for the top layer. This way, we realize individually addressable copies of our overlay target with steps in overlay error of \SI{37.5}{\nano\metre} in both X and Y directions, as visualized schematically in figure \ref{fig:method}c. Note that in the fabricated devices, the separation between individual overlay targets is more than ten times their width. We will refer to these synthetic overlay errors as overlay values, in order to avoid confusion with the error in measurement of these overlay values. One convenient property of this experimental architecture is that a small overlay error in our fabrication process will simply shift the origin of our overlay reference grid. In experiment, the overlay values in our overlay targets range from \SI{-187.5}{nm} to \SI{187.5}{nm} in $11$ steps along both X and Y, with uniform deviations of up to \SI{15}{\nano\metre} for separate arrays. This fabrication process deterministically provides us with targets with different overlay values, on which we may test overlay retrieval techniques.

\begin{figure}[htb!!]\centering
\includegraphics[width=.93\textwidth]{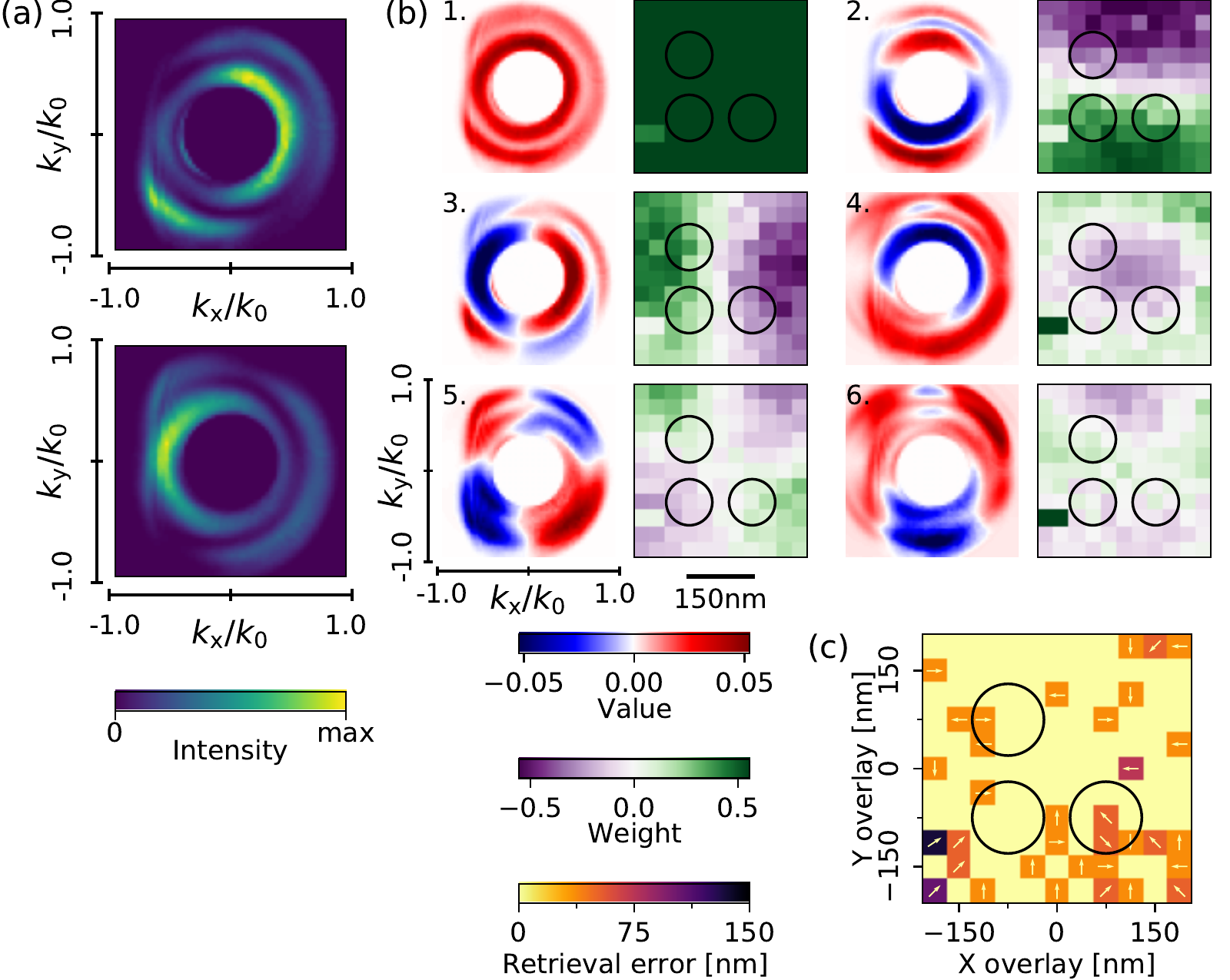}
\caption{(a) Example scattering patterns for overlay targets with overlay values \SI{-180}{\nano\metre} (top) and \SI{190}{\nano\metre} (bottom) along the X axis. The scattered light fills a cone of $\text{NA}_\text{out} = 0.95$. The dark region in the center corresponds to the illumination cone of $\text{NA}_\text{in} = 0.37$. (b) The first six elements of the singular value decomposition of a set of calibration data. Round, red-blue panels show the elements of the linear basis for angle-resolved intensity data and square, green-purple panels show the weight this element holds at each overlay value. Scale bars around element 5 show the extent in momentum space of the basis elements as well as the overlay range covered by the calibration. Black circles mark the size of the overlay target with respect to the calibration area; if these are taken as static, overlay value corresponds to the position of the fourth particle. (c) Retrieving overlay with the calibration data. Overlay retrieval error is calculated for new measurement data, using earlier calibration data from the same device. Where overlay value is not retrieved correctly, arrows indicate in which direction the error is made. Black circles show the scale of the overlay target, as in (b).} \label{fig:single}
\end{figure}

In order to determine overlay value, we collect angle-resolved intensity data. We illuminate the targets with supercontinuum laser light (NKT Whitelase Micro) filtered down to a \SI{10}{\nano\metre} band around \SI{620}{\nano\metre}. The effective numerical aperture of the illuminating beam is $\text{NA}_\text{in} = 0.37$, which produces a focal spot significantly wider than the overlay targets. This reduces the sensitivity of the experiment on the nanometer-scale positioning of the overlay target with respect to the laser spot. We collect the back-scattered light over a numerical aperture of $\text{NA}_\text{out} = 0.95$ through the same objective. The angle-resolved intensity distribution is projected onto a camera, as shown schematically in figure \ref{fig:method}d. We use a standard CMOS camera (Basler acA1920-40um) at an integration time of \SI{40}{\milli\second}, which is comparable with common practice in overlay metrology \cite{denBoef2016}.
We disregard the central $\text{NA}_\text{in} = 0.37$ that corresponds to specular reflection, instead looking at the light scattered out of this cone by the overlay targets. Two of the
scattering patterns thus collected are shown in figure \ref{fig:single}a. The images correspond to overlay values of around \SI{-180}{\nano\metre} (top) and \SI{190}{\nano\metre} (bottom) along the X axis. We see that the scattering patterns are clearly different and show beaming along the axis between different particles, in line with the literature on plasmonic phased array experiments \cite{Li2007, Koenderink2009}. These scattering patterns are to be mapped to overlay value.

We retrieve overlay value from angle-resolved intensity data by a calibration-based method previously used to retrieve the position of light sources \cite{Buijs2020} and scattering objects \cite{Wolterink2021}. By this method, we collect scattering patterns for all the overlay values available in the moir\'e pattern. We then use singular value decomposition, a tool from principal component analysis, to identify patterns in the relation between overlay value and scattering pattern. Singular value decomposition finds a linear basis for the set of intensity images, such that the total weight of each successive element in the basis is maximal \cite{Jolliffe2002}. For many problems, this permits compression of the reference dataset, where only the first few, largest-weight elements are preserved with minimal loss in information. In figure \ref{fig:single}b, we show the first six angle-resolved linear basis elements (round patterns, red-blue) and the corresponding weight each element has at the different overlay values (square patterns, green-purple). The bottom-layer nanoparticles are drawn for scale. From this perspective, overlay value corresponds to the position of the fourth, top-layer nanoparticle. After the first element, which corresponds to a largely overlay-independent background, we see two basis elements that map almost directly to Y and X overlay value. The latter, the third element, also explains the patterns we saw in figure \ref{fig:single}a. Further elements encode finer overlay value dependence of scattering patterns.
The optimal basis also provides some insight into how these overlay targets work, i.e. to what degree their response is due to near-field interactions between the nanoparticles versus far-field interference by the individual scattering signals of each nanoparticle. Near-field interactions are strongest when particles are close together, with a sharp position dependence. If near-field interactions were the dominant mechanism, we would likewise expect to see a sharp overlay value dependence in the radiation patterns, probably around the points where the top-layer particle sits directly over any of the bottom-layer particles. Instead, the dominant basis elements are nearly linear over the full \SI{350}{\nano\metre} overlay value range.
This suggests that the overlay target produces its scattering patterns mostly by single scattering and far-field interference.
Calculation of the optimal basis in figure \ref{fig:single}b constitutes calibration of the overlay target: we now know the overlay value dependence of scattering patterns on this set of devices.

In the first test, we collect the new scattering patterns on the same structures as were used for calibration. Experimental imperfections due to detection noise and optical re-alignment are thus not shared between both data-sets, but fabrication imperfections are. The new patterns are projected onto the optimal basis and we compare the coefficients with the weights found for every overlay value in the reference set. Summing the square of residuals gives a match value at each overlay value. The overlay value corresponding to the largest match is taken as our best estimate of overlay value given the measurement data and the calibration. We perform this procedure at every overlay value and calculate the retrieval error: the overlay value difference between the best estimate and the actual known overlay value. The results of this analysis are shown in figure \ref{fig:single}c. Arrows indicate in which direction the correct overlay value was missed for those cases where it was not retrieved correctly. We see that most overlay values are retrieved correctly, with the size of non-zero retrieval errors mostly a single \SI{37.5}{\nano\metre} step. We can further define an ensemble error $\Delta\text{OV}_\text{e}$, i.e. the average retrieval error across the entire calibrated domain: 
\begin{align}
\Delta\text{OV}_{\text{e}} = \frac{1}{n_x n_y} \sum_{(x,y)} |v_{xy}|
\end{align}
with $v_{xy}$ the individual retrieval error vectors for all overlay values $(x,y)$ on the $n_x \times n_y$ domain.
For these measurements, ensemble error is around $\Delta\text{OV}_{\text{e}} = \SI{20}{\nano\metre}$. We note that these errors are larger than the few-nanometer requirements in advanced nodes. However, this should be considered in the light of the enormous reduction in size, from \SI{100}{\micro\metre} to \SI{400}{\nano\metre} to a side. Although here we show just one realization, the errors in figure \ref{fig:single} do not appear to have a preferred direction, suggesting that combining multiple such sensors could improve performance. It is also conceivable that finer steps in the calibration set would reduce ensemble error. Further investigations might explore whether the source of the error lies in sensitivity to defects, positioning or something else and thus what the practical bounds on retrieval error are. From the available data, we conclude that overlay retrieval with the designed targets works reliably over the \SI{375}{\nano\metre} range.

\begin{figure}[htb!]\centering
\includegraphics[width=\textwidth]{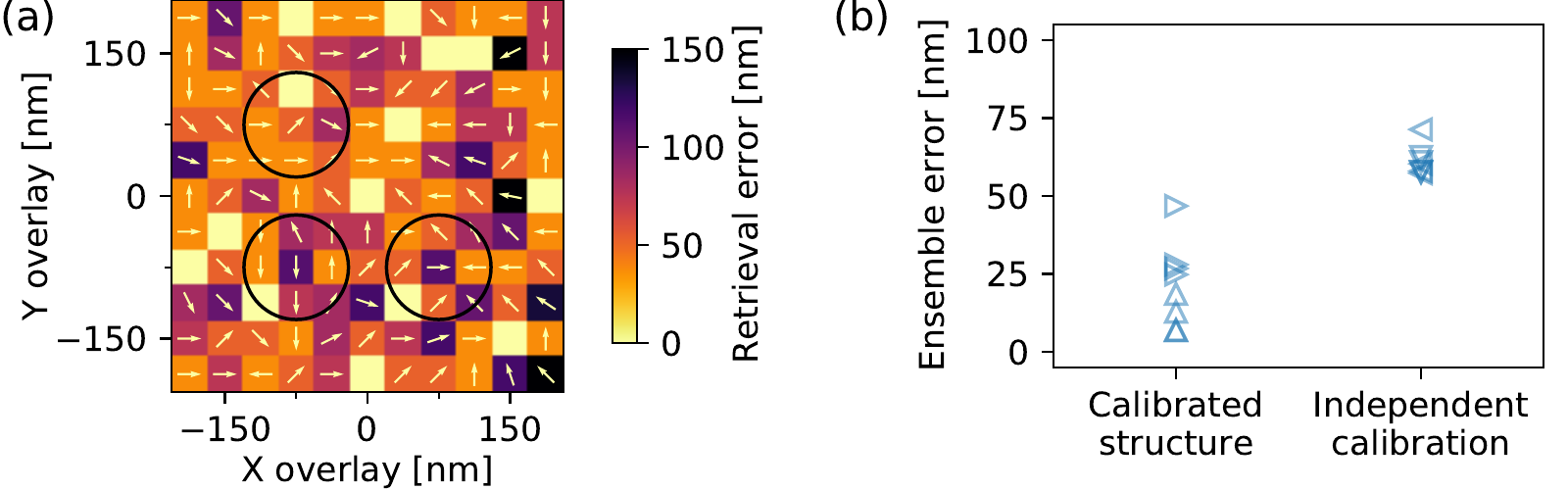}
\caption{(a) Overlay retrieval with independent calibration data, shown like figure \ref{fig:single}c. (b) Overlay retrieval performance across measurements, either using a fully calibrated device or retrieving overlay through independent calibration data. Differently oriented triangles correspond to different polarizations.} \label{fig:cross}
\end{figure}

So far, we have used devices that we had taken calibration data on beforehand. Of course, in overlay metrology, the target is specifically unknown, having been shaped by process variations that it is the purpose of metrology to calibrate. To test this more realistic case, we now reuse the optical calibration data from one device to retrieve overlay value on a separate one with its own fabrication imperfections. Analysing these in the same way as before, we find the results in figure \ref{fig:cross}a. We see that retrieval errors are appreciably larger than with a calibrated device, with ensemble error now around $\Delta\text{OV}_{\text{e}} = \SI{60}{\nano\metre}$. This increase in error is likely due to variability in our fabrication process. Indeed, a main challenge for benchmarking metrology innovations for CMOS technology using prototypes fabricated by e-beam technology, is that e-beam lithography does not reach the reliable fabrication quality of CMOS processing. Due to the small size of the targets, which leaves them vulnerable to nanometer-scale defects, a highly optimized process would be required to ensure reproducibility. To get a better feel for the source of errors, we have repeated our previous measurements, both calibrated on the same device and independently calibrated, on both devices, with different polarizations. We can also treat sample data as calibration data and vice versa. We show all ensemble errors found in this way in figure \ref{fig:cross}b. We see that the calibrated structure consistently outperforms the independently calibrated one, but also that the independently calibrated device consistently obtains an ensemble error around $\Delta\text{OV}_{\text{e}} = \SI{65}{\nano\metre}$. This suggest  that the differences between structures may cause a more or less fixed ensemble error level, rather than a wide spread. Of course, measurements on more devices of this type would be needed to further inform our interpretations and verify how non-ideal device fabrication affects overlay retrieval performance.

\section{Discussion}
We have performed two-dimensional overlay retrieval with tiny overlay targets, consisting of just four nanoparticles. Using a moir\'e grid to deterministically produce a wide range of overlay errors on a single chip, we show overlay retrieval error with mean error around \SI{20}{\nano\metre} on calibrated devices, using tens of millisecond integration times. This performance is obtained by principal component analysis of angle-resolved intensity data, which shows that the main variation in patterns maps nearly linearly to the two spatial axes. Reusing calibration data on separate structures shows overlay retrieval errors around \SI{65}{\nano\metre}, which is clearly larger than on calibrated structures, but nonetheless shows that the unavoidable defects arising from nanoparticle fabrication may be tolerable in overlay metrology.

The overlay targets demonstrate here are much smaller than typical overlay targets. Of course, larger overlay targets do provide stronger scattering signal, besides the added benefit of diffraction orders, which will affect the accuracy with which overlay error can be measured. This may be part of the reason why the current demonstration does not meet the commercial few-nanometer requirement. For the measurements with independent calibration, another reason may be sought in fabrication errors. The fabrication process employed is known to produce errors in particle size and shape that cause variance in scattering properties \cite{Karst2018}. We fully expect that the carefully controlled fabrication processes used in the semiconductor industry will drastically reduce variability between devices, bringing the ensemble error level for structures with independent calibration much closer to the error level of the fully calibrated devices. If the errors for calibrated structures can be shown to be uncorrelated, as seems to be the case from the data presented here, multiple minimal-footprint targets could be used to bring effective ensemble error down to tolerable levels. With the five order-of-magnitude difference in footprint, this would still represent a major gain in area. Our algorithm exclusively used angle-resolved intensity data to retrieve overlay. It is worth asking how the performance of this readout scheme compares to use of spectral information, as performed elsewhere \cite{Hentschel2012}, or use of other far-field degrees of freedom like polarization. Optimal far-field measurement schemes remain a fascinating target for further research.

Our minimal-footprint targets seem to encode overlay error in radiation patterns through far-field interference, rather than through more strongly position-dependent multiple scattering mechanisms. With this in mind, it is interesting to ask how more precise overlay sensors may be designed for a given measurement \cite{Adel2004, Roehrich2020}. One path could be to exploit phenomena with a stronger position dependence, such as plasmonic coupling between individual nanoparticles \cite{Soennichsen2005} or molecular ruler-like systems \cite{Weiss1999}. Overlay sensors are typically made, by necessity, out of the same material as the layers between which overlay error is to be measured. This means that metallic particles, as used here, can only be used in part of overlay applications. For others, dielectric (gap) resonators \cite{Kivshar2018} may be a good alternative, especially seeing how our technique does not appear to require multiple scattering between overlay target elements. As device features continue to shrink and overlay metrology becomes increasingly vital to process yield, we expect that innovation in overlay targets will prove critical.

\begin{acknowledgements}
The authors thank Wim Coene for valuable discussions.
This work is part of the research program of the Netherlands Organization for Scientific Research (NWO). It is part of the program High Tech Systems and Materials (HTSM) with Project No. 14669, which is (partly) financed by NWO.
\end{acknowledgements}

\bibliography{Wolterink2021APL.bib}

\begin{thebibliography}{22}%
\makeatletter
\providecommand \@ifxundefined [1]{%
 \@ifx{#1\undefined}
}%
\providecommand \@ifnum [1]{%
 \ifnum #1\expandafter \@firstoftwo
 \else \expandafter \@secondoftwo
 \fi
}%
\providecommand \@ifx [1]{%
 \ifx #1\expandafter \@firstoftwo
 \else \expandafter \@secondoftwo
 \fi
}%
\providecommand \natexlab [1]{#1}%
\providecommand \enquote  [1]{``#1''}%
\providecommand \bibnamefont  [1]{#1}%
\providecommand \bibfnamefont [1]{#1}%
\providecommand \citenamefont [1]{#1}%
\providecommand \href@noop [0]{\@secondoftwo}%
\providecommand \href [0]{\begingroup \@sanitize@url \@href}%
\providecommand \@href[1]{\@@startlink{#1}\@@href}%
\providecommand \@@href[1]{\endgroup#1\@@endlink}%
\providecommand \@sanitize@url [0]{\catcode `\\12\catcode `\$12\catcode
  `\&12\catcode `\#12\catcode `\^12\catcode `\_12\catcode `\%12\relax}%
\providecommand \@@startlink[1]{}%
\providecommand \@@endlink[0]{}%
\providecommand \url  [0]{\begingroup\@sanitize@url \@url }%
\providecommand \@url [1]{\endgroup\@href {#1}{\urlprefix }}%
\providecommand \urlprefix  [0]{URL }%
\providecommand \Eprint [0]{\href }%
\providecommand \doibase [0]{http://dx.doi.org/}%
\providecommand \selectlanguage [0]{\@gobble}%
\providecommand \bibinfo  [0]{\@secondoftwo}%
\providecommand \bibfield  [0]{\@secondoftwo}%
\providecommand \translation [1]{[#1]}%
\providecommand \BibitemOpen [0]{}%
\providecommand \bibitemStop [0]{}%
\providecommand \bibitemNoStop [0]{.\EOS\space}%
\providecommand \EOS [0]{\spacefactor3000\relax}%
\providecommand \BibitemShut  [1]{\csname bibitem#1\endcsname}%
\let\auto@bib@innerbib\@empty
\bibitem [{\citenamefont {Silver}, \citenamefont {Potzick},\ and\ \citenamefont
  {Larrabee}(1995)}]{Silver1995}%
  \BibitemOpen
  \bibfield  {author} {\bibinfo {author} {\bibfnamefont {R.~M.}\ \bibnamefont
  {Silver}}, \bibinfo {author} {\bibfnamefont {J.~E.}\ \bibnamefont {Potzick}},
  \ and\ \bibinfo {author} {\bibfnamefont {R.~D.}\ \bibnamefont {Larrabee}},\
  }in\ \href {\doibase 10.1117/12.209209} {\emph {\bibinfo {booktitle}
  {Integrated Circuit Metrology, Inspection, and Process Control {IX}}}},\
  \bibinfo {editor} {edited by\ \bibinfo {editor} {\bibfnamefont {M.~H.}\
  \bibnamefont {Bennett}}}\ (\bibinfo  {publisher} {{SPIE}},\ \bibinfo {year}
  {1995})\BibitemShut {NoStop}%
\bibitem [{\citenamefont {Bunday}(2016)}]{Bunday2016}%
  \BibitemOpen
  \bibfield  {author} {\bibinfo {author} {\bibfnamefont {B.}~\bibnamefont
  {Bunday}},\ }in\ \href {\doibase 10.1117/12.2218375} {\emph {\bibinfo
  {booktitle} {Metrology, Inspection, and Process Control for Microlithography
  {XXX}}}},\ \bibinfo {editor} {edited by\ \bibinfo {editor} {\bibfnamefont
  {M.~I.}\ \bibnamefont {Sanchez}}\ and\ \bibinfo {editor} {\bibfnamefont
  {V.~A.}\ \bibnamefont {Ukraintsev}}}\ (\bibinfo  {publisher} {{SPIE}},\
  \bibinfo {year} {2016})\BibitemShut {NoStop}%
\bibitem [{\citenamefont {Felix}\ \emph {et~al.}(2011)\citenamefont {Felix},
  \citenamefont {Gabor}, \citenamefont {Menon}, \citenamefont {Longo},
  \citenamefont {Halle}, \citenamefont {seng Koay},\ and\ \citenamefont
  {Colburn}}]{Felix2011}%
  \BibitemOpen
  \bibfield  {author} {\bibinfo {author} {\bibfnamefont {N.~M.}\ \bibnamefont
  {Felix}}, \bibinfo {author} {\bibfnamefont {A.~H.}\ \bibnamefont {Gabor}},
  \bibinfo {author} {\bibfnamefont {V.~C.}\ \bibnamefont {Menon}}, \bibinfo
  {author} {\bibfnamefont {P.~P.}\ \bibnamefont {Longo}}, \bibinfo {author}
  {\bibfnamefont {S.~D.}\ \bibnamefont {Halle}}, \bibinfo {author}
  {\bibfnamefont {C.}~\bibnamefont {seng Koay}}, \ and\ \bibinfo {author}
  {\bibfnamefont {M.~E.}\ \bibnamefont {Colburn}},\ }in\ \href {\doibase
  10.1117/12.879532} {\emph {\bibinfo {booktitle} {Metrology, Inspection, and
  Process Control for Microlithography {XXV}}}},\ \bibinfo {editor} {edited by\
  \bibinfo {editor} {\bibfnamefont {C.~J.}\ \bibnamefont {Raymond}}}\ (\bibinfo
   {publisher} {{SPIE}},\ \bibinfo {year} {2011})\BibitemShut {NoStop}%
\bibitem [{\citenamefont {Bunday}\ \emph {et~al.}(2018)\citenamefont {Bunday},
  \citenamefont {Bello}, \citenamefont {Solecky},\ and\ \citenamefont
  {Vaid}}]{Bunday2018}%
  \BibitemOpen
  \bibfield  {author} {\bibinfo {author} {\bibfnamefont {B.~D.}\ \bibnamefont
  {Bunday}}, \bibinfo {author} {\bibfnamefont {A.}~\bibnamefont {Bello}},
  \bibinfo {author} {\bibfnamefont {E.}~\bibnamefont {Solecky}}, \ and\
  \bibinfo {author} {\bibfnamefont {A.}~\bibnamefont {Vaid}},\ }in\ \href
  {\doibase 10.1117/12.2296679} {\emph {\bibinfo {booktitle} {Metrology,
  Inspection, and Process Control for Microlithography {XXXII}}}},\ \bibinfo
  {editor} {edited by\ \bibinfo {editor} {\bibfnamefont {O.}~\bibnamefont
  {Adan}}\ and\ \bibinfo {editor} {\bibfnamefont {V.~A.}\ \bibnamefont
  {Ukraintsev}}}\ (\bibinfo  {publisher} {{SPIE}},\ \bibinfo {year}
  {2018})\BibitemShut {NoStop}%
\bibitem [{\citenamefont {Orji}\ \emph {et~al.}(2018)\citenamefont {Orji},
  \citenamefont {Badaroglu}, \citenamefont {Barnes}, \citenamefont {Beitia},
  \citenamefont {Bunday}, \citenamefont {Celano}, \citenamefont {Kline},
  \citenamefont {Neisser}, \citenamefont {Obeng},\ and\ \citenamefont
  {Vladar}}]{Orji2018}%
  \BibitemOpen
  \bibfield  {author} {\bibinfo {author} {\bibfnamefont {N.~G.}\ \bibnamefont
  {Orji}}, \bibinfo {author} {\bibfnamefont {M.}~\bibnamefont {Badaroglu}},
  \bibinfo {author} {\bibfnamefont {B.~M.}\ \bibnamefont {Barnes}}, \bibinfo
  {author} {\bibfnamefont {C.}~\bibnamefont {Beitia}}, \bibinfo {author}
  {\bibfnamefont {B.~D.}\ \bibnamefont {Bunday}}, \bibinfo {author}
  {\bibfnamefont {U.}~\bibnamefont {Celano}}, \bibinfo {author} {\bibfnamefont
  {R.~J.}\ \bibnamefont {Kline}}, \bibinfo {author} {\bibfnamefont
  {M.}~\bibnamefont {Neisser}}, \bibinfo {author} {\bibfnamefont
  {Y.}~\bibnamefont {Obeng}}, \ and\ \bibinfo {author} {\bibfnamefont {A.~E.}\
  \bibnamefont {Vladar}},\ }\href {\doibase 10.1038/s41928-018-0150-9}
  {\bibfield  {journal} {\bibinfo  {journal} {Nat. Electron.}\ }\textbf
  {\bibinfo {volume} {1}},\ \bibinfo {pages} {532} (\bibinfo {year}
  {2018})}\BibitemShut {NoStop}%
\bibitem [{\citenamefont {den Boef}(2016)}]{denBoef2016}%
  \BibitemOpen
  \bibfield  {author} {\bibinfo {author} {\bibfnamefont {A.~J.}\ \bibnamefont
  {den Boef}},\ }\href {\doibase 10.1088/2051-672x/4/2/023001} {\bibfield
  {journal} {\bibinfo  {journal} {Surf. Topogr.: Metrol. Prop.}\ }\textbf
  {\bibinfo {volume} {4}},\ \bibinfo {pages} {023001} (\bibinfo {year}
  {2016})}\BibitemShut {NoStop}%
\bibitem [{\citenamefont {Bishop}\ \emph {et~al.}(1991)\citenamefont {Bishop},
  \citenamefont {Gaspar}, \citenamefont {Milner}, \citenamefont {Naqvi},\ and\
  \citenamefont {McNeil}}]{Bishop1991}%
  \BibitemOpen
  \bibfield  {author} {\bibinfo {author} {\bibfnamefont {K.~P.}\ \bibnamefont
  {Bishop}}, \bibinfo {author} {\bibfnamefont {S.~M.}\ \bibnamefont {Gaspar}},
  \bibinfo {author} {\bibfnamefont {L.-M.}\ \bibnamefont {Milner}}, \bibinfo
  {author} {\bibfnamefont {S.~S.~H.}\ \bibnamefont {Naqvi}}, \ and\ \bibinfo
  {author} {\bibfnamefont {J.~R.}\ \bibnamefont {McNeil}},\ }in\ \href
  {\doibase 10.1117/12.49402} {\emph {\bibinfo {booktitle} {International
  Conference on the Application and Theory of Periodic Structures}}},\ \bibinfo
  {editor} {edited by\ \bibinfo {editor} {\bibfnamefont {J.~M.}\ \bibnamefont
  {Lerner}}\ and\ \bibinfo {editor} {\bibfnamefont {W.~R.}\ \bibnamefont
  {McKinney}}}\ (\bibinfo  {publisher} {{SPIE}},\ \bibinfo {year}
  {1991})\BibitemShut {NoStop}%
\bibitem [{\citenamefont {Bouwhuis}\ and\ \citenamefont
  {Wittekoek}(1979)}]{Bouwhuis1979}%
  \BibitemOpen
  \bibfield  {author} {\bibinfo {author} {\bibfnamefont {G.}~\bibnamefont
  {Bouwhuis}}\ and\ \bibinfo {author} {\bibfnamefont {S.}~\bibnamefont
  {Wittekoek}},\ }\href {\doibase 10.1109/t-ed.1979.19483} {\bibfield
  {journal} {\bibinfo  {journal} {{IEEE} Trans. Electron Devices}\ }\textbf
  {\bibinfo {volume} {26}},\ \bibinfo {pages} {723} (\bibinfo {year}
  {1979})}\BibitemShut {NoStop}%
\bibitem [{\citenamefont {van Dijk}\ \emph {et~al.}(2006)\citenamefont {van
  Dijk}, \citenamefont {Tchebotareva}, \citenamefont {Orrit}, \citenamefont
  {Lippitz}, \citenamefont {Berciaud}, \citenamefont {Lasne}, \citenamefont
  {Cognet},\ and\ \citenamefont {Lounis}}]{vanDijk2006}%
  \BibitemOpen
  \bibfield  {author} {\bibinfo {author} {\bibfnamefont {M.~A.}\ \bibnamefont
  {van Dijk}}, \bibinfo {author} {\bibfnamefont {A.~L.}\ \bibnamefont
  {Tchebotareva}}, \bibinfo {author} {\bibfnamefont {M.}~\bibnamefont {Orrit}},
  \bibinfo {author} {\bibfnamefont {M.}~\bibnamefont {Lippitz}}, \bibinfo
  {author} {\bibfnamefont {S.}~\bibnamefont {Berciaud}}, \bibinfo {author}
  {\bibfnamefont {D.}~\bibnamefont {Lasne}}, \bibinfo {author} {\bibfnamefont
  {L.}~\bibnamefont {Cognet}}, \ and\ \bibinfo {author} {\bibfnamefont
  {B.}~\bibnamefont {Lounis}},\ }\href {\doibase 10.1039/b606090k} {\bibfield
  {journal} {\bibinfo  {journal} {Phys. Chem. Chem. Phys.}\ }\textbf {\bibinfo
  {volume} {8}},\ \bibinfo {pages} {3486} (\bibinfo {year} {2006})}\BibitemShut
  {NoStop}%
\bibitem [{\citenamefont {Weber}\ and\ \citenamefont {Ford}(2004)}]{Weber2004}%
  \BibitemOpen
  \bibfield  {author} {\bibinfo {author} {\bibfnamefont {W.~H.}\ \bibnamefont
  {Weber}}\ and\ \bibinfo {author} {\bibfnamefont {G.~W.}\ \bibnamefont
  {Ford}},\ }\href {\doibase 10.1103/physrevb.70.125429} {\bibfield  {journal}
  {\bibinfo  {journal} {Phys. Rev. B}\ }\textbf {\bibinfo {volume} {70}},\
  \bibinfo {pages} {125429} (\bibinfo {year} {2004})}\BibitemShut {NoStop}%
\bibitem [{\citenamefont {Kivshar}(2018)}]{Kivshar2018}%
  \BibitemOpen
  \bibfield  {author} {\bibinfo {author} {\bibfnamefont {Y.}~\bibnamefont
  {Kivshar}},\ }\href {\doibase 10.1093/nsr/nwy017} {\bibfield  {journal}
  {\bibinfo  {journal} {Natl. Sci. Rev.}\ }\textbf {\bibinfo {volume} {5}},\
  \bibinfo {pages} {144} (\bibinfo {year} {2018})}\BibitemShut {NoStop}%
\bibitem [{\citenamefont {Hentschel}\ \emph {et~al.}(2012)\citenamefont
  {Hentschel}, \citenamefont {Sch\"{a}ferling}, \citenamefont {Weiss},
  \citenamefont {Liu},\ and\ \citenamefont {Giessen}}]{Hentschel2012}%
  \BibitemOpen
  \bibfield  {author} {\bibinfo {author} {\bibfnamefont {M.}~\bibnamefont
  {Hentschel}}, \bibinfo {author} {\bibfnamefont {M.}~\bibnamefont
  {Sch\"{a}ferling}}, \bibinfo {author} {\bibfnamefont {T.}~\bibnamefont
  {Weiss}}, \bibinfo {author} {\bibfnamefont {N.}~\bibnamefont {Liu}}, \ and\
  \bibinfo {author} {\bibfnamefont {H.}~\bibnamefont {Giessen}},\ }\href
  {\doibase 10.1021/nl300769x} {\bibfield  {journal} {\bibinfo  {journal} {Nano
  Lett.}\ }\textbf {\bibinfo {volume} {12}},\ \bibinfo {pages} {2542} (\bibinfo
  {year} {2012})}\BibitemShut {NoStop}%
\bibitem [{\citenamefont {Li}, \citenamefont {Salandrino},\ and\ \citenamefont
  {Engheta}(2007)}]{Li2007}%
  \BibitemOpen
  \bibfield  {author} {\bibinfo {author} {\bibfnamefont {J.}~\bibnamefont
  {Li}}, \bibinfo {author} {\bibfnamefont {A.}~\bibnamefont {Salandrino}}, \
  and\ \bibinfo {author} {\bibfnamefont {N.}~\bibnamefont {Engheta}},\ }\href
  {\doibase 10.1103/physrevb.76.245403} {\bibfield  {journal} {\bibinfo
  {journal} {Phys. Rev. B}\ }\textbf {\bibinfo {volume} {76}},\ \bibinfo
  {pages} {245403} (\bibinfo {year} {2007})}\BibitemShut {NoStop}%
\bibitem [{\citenamefont {Koenderink}(2009)}]{Koenderink2009}%
  \BibitemOpen
  \bibfield  {author} {\bibinfo {author} {\bibfnamefont {A.~F.}\ \bibnamefont
  {Koenderink}},\ }\href {\doibase 10.1021/nl902439n} {\bibfield  {journal}
  {\bibinfo  {journal} {Nano Lett.}\ }\textbf {\bibinfo {volume} {9}},\
  \bibinfo {pages} {4228} (\bibinfo {year} {2009})}\BibitemShut {NoStop}%
\bibitem [{\citenamefont {Buijs}\ \emph {et~al.}(2020)\citenamefont {Buijs},
  \citenamefont {Schilder}, \citenamefont {Wolterink}, \citenamefont {Gerini},
  \citenamefont {Verhagen},\ and\ \citenamefont {Koenderink}}]{Buijs2020}%
  \BibitemOpen
  \bibfield  {author} {\bibinfo {author} {\bibfnamefont {R.~D.}\ \bibnamefont
  {Buijs}}, \bibinfo {author} {\bibfnamefont {N.~J.}\ \bibnamefont {Schilder}},
  \bibinfo {author} {\bibfnamefont {T.~A.~W.}\ \bibnamefont {Wolterink}},
  \bibinfo {author} {\bibfnamefont {G.}~\bibnamefont {Gerini}}, \bibinfo
  {author} {\bibfnamefont {E.}~\bibnamefont {Verhagen}}, \ and\ \bibinfo
  {author} {\bibfnamefont {A.~F.}\ \bibnamefont {Koenderink}},\ }\href
  {\doibase 10.1021/acsphotonics.0c01350} {\bibfield  {journal} {\bibinfo
  {journal} {{ACS} Photonics}\ }\textbf {\bibinfo {volume} {7}},\ \bibinfo
  {pages} {3246} (\bibinfo {year} {2020})}\BibitemShut {NoStop}%
\bibitem [{\citenamefont {Wolterink}\ \emph {et~al.}(2021)\citenamefont
  {Wolterink}, \citenamefont {Buijs}, \citenamefont {Gerini}, \citenamefont
  {Koenderink},\ and\ \citenamefont {Verhagen}}]{Wolterink2021}%
  \BibitemOpen
  \bibfield  {author} {\bibinfo {author} {\bibfnamefont {T.~A.~W.}\
  \bibnamefont {Wolterink}}, \bibinfo {author} {\bibfnamefont {R.~D.}\
  \bibnamefont {Buijs}}, \bibinfo {author} {\bibfnamefont {G.}~\bibnamefont
  {Gerini}}, \bibinfo {author} {\bibfnamefont {A.~F.}\ \bibnamefont
  {Koenderink}}, \ and\ \bibinfo {author} {\bibfnamefont {E.}~\bibnamefont
  {Verhagen}},\ }\href {\doibase 10.1515/nanoph-2020-0669} {\bibfield
  {journal} {\bibinfo  {journal} {Nanophotonics}\ }\textbf {\bibinfo {volume}
  {10}},\ \bibinfo {pages} {1723} (\bibinfo {year} {2021})}\BibitemShut
  {NoStop}%
\bibitem [{\citenamefont {Jolliffe}(2002)}]{Jolliffe2002}%
  \BibitemOpen
  \bibfield  {author} {\bibinfo {author} {\bibfnamefont {I.~T.}\ \bibnamefont
  {Jolliffe}},\ }\href {\doibase 10.1007/b98835} {\emph {\bibinfo {title}
  {Principal Component Analysis}}}\ (\bibinfo  {publisher} {Springer-Verlag},\
  \bibinfo {year} {2002})\BibitemShut {NoStop}%
\bibitem [{\citenamefont {Karst}\ \emph {et~al.}(2018)\citenamefont {Karst},
  \citenamefont {Strohfeldt}, \citenamefont {Sch\"{a}ferling}, \citenamefont
  {Giessen},\ and\ \citenamefont {Hentschel}}]{Karst2018}%
  \BibitemOpen
  \bibfield  {author} {\bibinfo {author} {\bibfnamefont {J.}~\bibnamefont
  {Karst}}, \bibinfo {author} {\bibfnamefont {N.}~\bibnamefont {Strohfeldt}},
  \bibinfo {author} {\bibfnamefont {M.}~\bibnamefont {Sch\"{a}ferling}},
  \bibinfo {author} {\bibfnamefont {H.}~\bibnamefont {Giessen}}, \ and\
  \bibinfo {author} {\bibfnamefont {M.}~\bibnamefont {Hentschel}},\ }\href
  {\doibase 10.1002/adom.201800087} {\bibfield  {journal} {\bibinfo  {journal}
  {Adv. Opt. Mater.}\ }\textbf {\bibinfo {volume} {6}},\ \bibinfo {pages}
  {1800087} (\bibinfo {year} {2018})}\BibitemShut {NoStop}%
\bibitem [{\citenamefont {Adel}\ \emph {et~al.}(2004)\citenamefont {Adel},
  \citenamefont {Ghinovker}, \citenamefont {Golovanevsky}, \citenamefont
  {Izikson}, \citenamefont {Kassel}, \citenamefont {Yaffe}, \citenamefont
  {Bruckstein}, \citenamefont {Goldenberg}, \citenamefont {Rubner},\ and\
  \citenamefont {Rudzsky}}]{Adel2004}%
  \BibitemOpen
  \bibfield  {author} {\bibinfo {author} {\bibfnamefont {M.}~\bibnamefont
  {Adel}}, \bibinfo {author} {\bibfnamefont {M.}~\bibnamefont {Ghinovker}},
  \bibinfo {author} {\bibfnamefont {B.}~\bibnamefont {Golovanevsky}}, \bibinfo
  {author} {\bibfnamefont {P.}~\bibnamefont {Izikson}}, \bibinfo {author}
  {\bibfnamefont {E.}~\bibnamefont {Kassel}}, \bibinfo {author} {\bibfnamefont
  {D.}~\bibnamefont {Yaffe}}, \bibinfo {author} {\bibfnamefont
  {A.}~\bibnamefont {Bruckstein}}, \bibinfo {author} {\bibfnamefont
  {R.}~\bibnamefont {Goldenberg}}, \bibinfo {author} {\bibfnamefont
  {Y.}~\bibnamefont {Rubner}}, \ and\ \bibinfo {author} {\bibfnamefont
  {M.}~\bibnamefont {Rudzsky}},\ }\href {\doibase 10.1109/tsm.2004.826955}
  {\bibfield  {journal} {\bibinfo  {journal} {{IEEE} Trans. Semicond. Manuf.}\
  }\textbf {\bibinfo {volume} {17}},\ \bibinfo {pages} {166} (\bibinfo {year}
  {2004})}\BibitemShut {NoStop}%
\bibitem [{\citenamefont {R\"{o}hrich}\ \emph {et~al.}(2020)\citenamefont
  {R\"{o}hrich}, \citenamefont {Oliveri}, \citenamefont {Kovaios},
  \citenamefont {Tenner}, \citenamefont {den Boef}, \citenamefont {Overvelde},\
  and\ \citenamefont {Koenderink}}]{Roehrich2020}%
  \BibitemOpen
  \bibfield  {author} {\bibinfo {author} {\bibfnamefont {R.}~\bibnamefont
  {R\"{o}hrich}}, \bibinfo {author} {\bibfnamefont {G.}~\bibnamefont
  {Oliveri}}, \bibinfo {author} {\bibfnamefont {S.}~\bibnamefont {Kovaios}},
  \bibinfo {author} {\bibfnamefont {V.~T.}\ \bibnamefont {Tenner}}, \bibinfo
  {author} {\bibfnamefont {A.~J.}\ \bibnamefont {den Boef}}, \bibinfo {author}
  {\bibfnamefont {J.~T.~B.}\ \bibnamefont {Overvelde}}, \ and\ \bibinfo
  {author} {\bibfnamefont {A.~F.}\ \bibnamefont {Koenderink}},\ }\href
  {\doibase 10.1021/acsphotonics.0c00911} {\bibfield  {journal} {\bibinfo
  {journal} {{ACS} Photonics}\ }\textbf {\bibinfo {volume} {7}},\ \bibinfo
  {pages} {2765} (\bibinfo {year} {2020})}\BibitemShut {NoStop}%
\bibitem [{\citenamefont {S\"{o}nnichsen}\ \emph {et~al.}(2005)\citenamefont
  {S\"{o}nnichsen}, \citenamefont {Reinhard}, \citenamefont {Liphardt},\ and\
  \citenamefont {Alivisatos}}]{Soennichsen2005}%
  \BibitemOpen
  \bibfield  {author} {\bibinfo {author} {\bibfnamefont {C.}~\bibnamefont
  {S\"{o}nnichsen}}, \bibinfo {author} {\bibfnamefont {B.~M.}\ \bibnamefont
  {Reinhard}}, \bibinfo {author} {\bibfnamefont {J.}~\bibnamefont {Liphardt}},
  \ and\ \bibinfo {author} {\bibfnamefont {A.~P.}\ \bibnamefont {Alivisatos}},\
  }\href {\doibase 10.1038/nbt1100} {\bibfield  {journal} {\bibinfo  {journal}
  {Nat. Biotechnol.}\ }\textbf {\bibinfo {volume} {23}},\ \bibinfo {pages}
  {741} (\bibinfo {year} {2005})}\BibitemShut {NoStop}%
\bibitem [{\citenamefont {Weiss}(1999)}]{Weiss1999}%
  \BibitemOpen
  \bibfield  {author} {\bibinfo {author} {\bibfnamefont {S.}~\bibnamefont
  {Weiss}},\ }\href {\doibase 10.1126/science.283.5408.1676} {\bibfield
  {journal} {\bibinfo  {journal} {Science}\ }\textbf {\bibinfo {volume}
  {283}},\ \bibinfo {pages} {1676} (\bibinfo {year} {1999})}\BibitemShut
  {NoStop}%
\end{thebibliography}%

\end{document}